\documentclass[twocolumn,showpacs,preprintnumbers,nofootinbib,prd,superscriptaddress,groupedaddress,10pt]{revtex4-1}

\usepackage[centertags]{amsmath}
\usepackage{amssymb}
\usepackage{latexsym}
\usepackage{enumerate}
\usepackage{graphicx}
\usepackage{mathrsfs}
\usepackage{hyperref}
\usepackage{stmaryrd}
\usepackage{import}
\usepackage{tensor}
\usepackage{color}
\usepackage{bm}
\usepackage{multirow}
\usepackage{float}

\newcommand{\bi}{\begin{itemize}}
\newcommand{\ei}{\end{itemize}}
\newcommand{\be}{\begin{equation}}
\newcommand{\ee}{\end{equation}}

\renewcommand{\l}{\left(}
\renewcommand{\r}{\right)}

\renewcommand{\o}{\omega}
\renewcommand{\th}{\theta}

\newcommand{\q}{\quad}

\newcommand{\pa}{\partial}

\def\be{\begin{equation}}
\def\ee{\end{equation}}
\def\bea{\begin{eqnarray}}
\def\eea{\end{eqnarray}}

\newcommand{\beq}{\begin{eqnarray}}
\newcommand{\eeq}{\end{eqnarray}}

\begin{document}

\title{Scattering of point particles by black holes: gravitational radiation}

\author{
Seth Hopper$^{1}$,
Vitor Cardoso$^{1,2}$
}
\affiliation{${^1}$ CENTRA, Departamento de F\'{\i}sica, Instituto Superior T\'ecnico -- IST, Universidade de Lisboa -- UL,
Avenida Rovisco Pais 1, 1049 Lisboa, Portugal}
\affiliation{${^2}$ Perimeter Institute for Theoretical Physics, 31 Caroline Street North
Waterloo, Ontario N2L 2Y5, Canada}

\begin{abstract}
Gravitational waves can teach us
not only about sources and the environment where they were generated, but also
about the gravitational interaction itself. Here we study the features 
of gravitational radiation produced during the 
scattering of a point-like mass by
a black hole. Our results are exact (to numerical error) 
at any order in a velocity 
expansion, and are compared against various approximations.
At large impact parameter and relatively small velocities our results agree to
within percent level with various post-Newtonian and weak-field results.
Further, we find good agreement 
with scaling predictions in the weak-field/high-energy regime.
Lastly, we achieve striking agreement with zero-frequency estimates.
\end{abstract}

\maketitle

\section{Introduction}
\label{sec:intro}
The momentous direct detection of gravitational waves (GWs) by the LIGO collaboration 
~\cite{Abbott:2016blz,Abbott:2016nmj,AbboETC17} provided the strongest evidence to date
that black holes (BHs) exist, merge, and that the GWs 
produced in the strong-field regime are well described by Einstein's theory.
As more sensitive detectors come online, precise 
tests of strong-field gravity will become 
possible~\cite{Gair:2012nm,Yunes:2013dva,Berti:2015itd}. 
Deviations from the Kerr geometry
will be imprinted, for example, in the way that BHs 
`vibrate'~\cite{Berti:2005ys,Berti:2016lat}. Fundamental 
issues -- such as the existence of horizons in our universe 
cloaking spacetime singularities -- can finally be addressed by 
either looking at smoking-gun signs of surfaces 
(such as `echoes')~\cite{Cardoso:2016rao,Cardoso:2016oxy} 
or the way that the inspiral proceeds~\cite{Maselli:2017cmm,Sennett:2017etc}.
Recent research into the GWs produced by compact binaries 
has been dominated by the study of bound motion,
which is becoming quite well understood. Thus, we feel it is an
appropriate time to refocus attention on the unbound binary problem.

GWs are also the perfect tool to understand gravity at its extreme. 
High-energy collisions of BHs produce the largest 
luminosities in the universe, which are thought to saturate
the Dyson-Gibbons-Thorne bound ($c^5/G$) on this quantity~\cite{Cardoso:2013krh}.
High-energy collisions of {\it particles} are expected to be universal. 
At large enough center-of-mass energy, the end state is always a BH and the details
of the initial state are forever hidden behind its 
horizon~\cite{Choptuik:2009ww,Sperhake:2012me,Cardoso:2014uka}.
Recently, it was argued that the scattering of two particles at large enough energies
and small deflection angles has some interesting properties, that the spectrum
of graviton emission takes a simple and elegant limiting form, 
which can be used to learn about the S-matrix
of quantum gravity~\cite{Gruzinov:2014moa,Ciafaloni:2015vsa}.

With the above as motivation, we start here the investigation 
of scattering of particles by BHs. 
In this work we focus primarily on the low/medium-energy regime where
particle speeds (at infinite separation) do not exceed $0.25c$. 
We also briefly explore high-energy events with with speeds reaching $0.98c$.
In the future we plan to use the code that we have developed here to more-thoroughly
explore the high energy regime. 
There have been a number of approximate, analytic works done over the years.
Of particular interest are works 
by Peters \cite{Peters:1970mx}, Smarr \cite{Smarr:1977fy}, 
Kovacs and Thorne \cite{Kovacs:1978eu}, Turner \cite{Turn77},
and Blanchet and Sch\"afer \cite{Blanchet:1989cu}, all of which we compare to here.

The remaining of this paper is organized as follows. Sec.~\ref{sec:setup}
establishes notation regarding the physical system we are studying and presents the 
numerical method we use to solve the first-order Einstein equations.
Sec.~\ref{sec:results} presents known analytic predictions and our results
found by comparing with those expressions. Sec.~\ref{sec:conc} provides
brief concluding remarks. Throughout this paper, unless otherwise mentioned,
we use Schwarzschild coordinates $t,r,\theta, \phi$, and
a subscript $p$ indicates a field evaluated at the particle's location. 
We work in coordinates where $G = c = 1$, although 
we do explicitly include factors of $G$ and $c$
in some expressions when it adds clarity.

\section{Setup and numerical procedure}
\label{sec:setup}

\subsection{Scattering geodesics on Schwarzschild}
\label{sec:geos}
We consider point particle (mass $\mu$) motion on a Schwarzschild background
(mass $M$).
Take the worldline to be parametrized by proper time $\tau$, 
i.e.~$x_p^{\mu}(\tau)$, with associated four-velocity is $u^{\mu} = dx_p^{\mu}/d\tau$.
We confine the particle to $\th_p = \pi/2$ without loss of generality and hence write
$x^\mu_p(\tau)= \left[ t_p(\tau),r_p(\tau),\pi/2,\phi_p(\tau) \right]$.
Generic geodesics are best parametrized by the constants of motion,
$\mathcal{E}$ and $\mathcal{L}$, respectively, the specific energy and 
angular momentum. They relate to the four-velocity via
\be
\label{eq:fourVelocity}
u^t = \frac{\mathcal{E}}{f_{p}}\,,  
\q \q
u^{\varphi} = \frac{{\cal{L}}}{r_p^2} \,,
\q \q
\l u^r \r^2 = \mathcal{E}^2-U^2_{p}\,, 
\ee
where the effective potential is
\be
\label{eq:effV}
U^2(r,\mathcal{L}^2) \equiv f \l 1 + \frac{\mathcal{L}^2}{r^2} \r ,
\ee
and $f \equiv  1 - 2M /r$.

For scattering geodesics, it is convenient to replace $\mathcal{L}$
with either the periapsis $r_{\rm min}$ or the impact parameter $b$.
The former is related to $\mathcal{E}$ and $\mathcal{L}$ via
$U^2(r_{\rm min},\mathcal{L}^2) = \mathcal{E}^2$,
while the latter is defined as
$ b \equiv \mathcal{L} / \sqrt{\mathcal{E}^2 - 1 }.$
Note that $b \to \infty$ when $\mathcal{E} = 1$,
making $r_{\rm min}$ the preferable choice in the case of parabolic motion.

\subsection{The frequency domain, RWZ formalism for unbound motion}
We developed a new perturbation theory code to obtain the numerical
results presented in this work.
While the details of that code and its method will be discussed 
in an accompanying paper~\cite{Hopp17}, here we provide a brief summary.
We use the Regge-Wheeler-Zerilli (RWZ) \cite{ReggWhee57,Zeri70} formalism
wherein the field equations of
any radiative $\ell m$ mode reduce to a single 1+1 wave equation,
\begin{align}
\left[-\frac{\pa^2}{\pa t^2} + \frac{\pa^2}{\pa r_*^2} - V_\ell(r)\right]
\Psi_{\ell m}(t,r) = S_{\ell m}(t,r).
\label{eq:TDmastereqnSrc}
\end{align}
Here $r_*$ is the usual tortoise coordinate, $r_* = r + 2M \log (r/2M - 1)$.
The `master function' $\Psi_{\ell m}$, its source $S_{\ell m}$ 
and the potential  $V_\ell$ are all $\ell+m$ parity-dependent.
We choose to work in the frequency domain (FD),
and thus assume that the field and its source,
can be represented by integrals over Fourier harmonics,
\begin{align}
\label{eq:invFourierTransf}
\begin{split}
\Psi_{\ell m}(t,r) &= \frac{1}{2\pi} \int_{-\infty}^\infty 
X_{\ell m\o}(r) \, e^{-i \o t}  d\o \,,\\
S_{\ell m}(t,r) &= \frac{1}{2\pi} \int_{-\infty}^\infty 
Z_{\ell m\o}(r) \, e^{-i \o t}  d\o \,.
\end{split}
\end{align}
In the FD the master equation \eqref{eq:TDmastereqnSrc}
 takes on the following form
\begin{align}
\label{eq:FDmastereqn}
\left[\frac{d^2}{dr_*^2} +\o^2 -V_\ell(r)\right]
X_{\ell m\o}(r) = Z_{\ell m\o}(r).
\end{align}
At infinity and the horizon we take as boundary conditions
retarded, unit-amplitude, homogeneous traveling waves,
\begin{align}
\hat{X}^\pm_{\ell m\o}(r_* \to \pm \infty )
= e^{\pm i \omega r_*} .
\end{align}
The solution to Eq.~\eqref{eq:FDmastereqn} 
follows from the method of variation of parameters,
\begin{align}
\label{eq:FDInhomog}
X_{\ell m\o} (r) = c^+_{\ell m\o}(r) \, \hat{X}^+_{\ell m\o}(r)
+ c^-_{\ell m\o}(r) \, \hat{X}^-_{\ell m\o}(r) ,
\end{align}
where
\begin{align}
\label{eq:cPM}
\begin{split}
c^+_{\ell m\o}(r) &= \frac{1}{W_{\ell m\o}} \, \int_{2M}^{r} 
\frac{dr'}{f(r')} \ \hat{X}^-_{\ell m\o}(r') \, Z_{\ell m\o}(r') \,,\\
c^-_{\ell m\o}(r) &= \frac{1}{W_{\ell m\o}} \, \int_{r}^{\infty} 
\frac{dr'}{f(r')} \ \hat{X}^+_{\ell m\o}(r') \, Z_{\ell m\o}(r') \,,
\end{split}
\end{align}
and $W_{\ell m\o}$ is the (constant-in-$r_*$) Wronskian.
Extending the integrals in Eq.~\eqref{eq:cPM} over all space provides the 
\emph{normalization coefficients},
\begin{align}
\label{eq:normC}
C_{\ell m\o}^{\pm} 
= \frac{1}{W_{\ell m\o}} \int_{2M}^{\infty} \frac{dr}{f}
\ \hat X^{\mp}_{\ell m\o} (r) Z_{\ell m\o} (r).
\end{align}
These coefficients are all that one needs 
to compute radiated energy and waveforms, which we find in this work.
The details of solving Eq.~\eqref{eq:normC} (which makes up the brunt
of our numerical calculation) are involved. In particular, the FD source
term $Z_{\ell m \omega}$ depends on the particular choice of master function,
a choice which affects the convergence of the integral in Eq.~\eqref{eq:normC}. 
As mentioned, this will be covered in detail in an accompanying work.

Assuming we have solved Eq.~\eqref{eq:normC} for a range of harmonics,
we compute the waveform at retarded time $u = t - r_*$ and $r_* \to \infty$
via
\begin{align}
\label{eq:waveform}
\Psi_{\ell m}(u, r_*\to \infty) 
&= 	
\frac{1}{2\pi} \int_{-\infty}^\infty 
C^+_{\ell m\o} \, e^{-i \o u}  d\o .
\end{align}
This follows because, as $r_* \to \infty$ the FD particular solutions go to
$X^+_{\ell m \omega} (r_* \to \infty) = C^+_{\ell m\omega} e^{i \o r_*}$.
One can sum these over $\ell$ and $m$ 
to form the transverse-traceless metric perturbation,
as shown in Ref.~\cite{Martel:2005ir}.

Our code also provides flux and energy spectrum results.
When using the Zerilli-Moncrief \cite{Moncrief:1974am} (for $\ell+m$ even) and 
Cunningham-Price-Moncrief \cite{Cunningham:1978zfa} ($\ell+m$ odd) variables, 
the energy flux at infinity is
\be
\label{eq:EDot}
\dot E^+_{\ell m} (u, r_* \to \infty) = \frac{1}{64 \pi} 
\frac{(\ell+2)!}{(\ell-2)!}
\left| \dot \Psi^{+}_{\ell m} (u, r_* \to \infty) \right|^2.
\ee
Then, the total energy radiated for a given $\ell m$ mode to infinity is
\begin{align}
\label{eq:enRad}
 E^+_{\ell m}
= \frac{1}{128 \pi^2} 
\frac{(\ell+2)!}{(\ell-2)!}
\int \o^{2}  \left|C^\pm_{\ell m\o}\right|^2 d\o\, .
\end{align}
In practice, we discretize the integrals \eqref{eq:waveform} and \eqref{eq:enRad}
(the smallest frequencies our machine-precision code can reach are $\sim 10^{-6} / M$). 
Then, we add positive and negative harmonics until the total radiated energy 
converges to a relative error of 
at most 0.1\% (in practice we usually achieve a much
greater level of accuracy). As shown in later sections, our greatest limitation
is the small (in magnitude) frequencies, in particular, for systems
with very large $r_{\rm min}$.

For most of the results we compare to in this paper, we do not require
modes of $\ell > 6$. However, for the high-energy results given below in 
Sec.~\ref{sec:weakField},
we do need higher $\ell$ modes.
Given the modest accuracy requirements of this work, we are able to
truncate out $\ell$-sum at 8 at which point we can clearly see
the exponential convergence and fit out higher-order contributions
to the sum. This saves computational time and gives results accurate
to better than 1\%.


\begin{figure*}
\includegraphics[width=\textwidth]{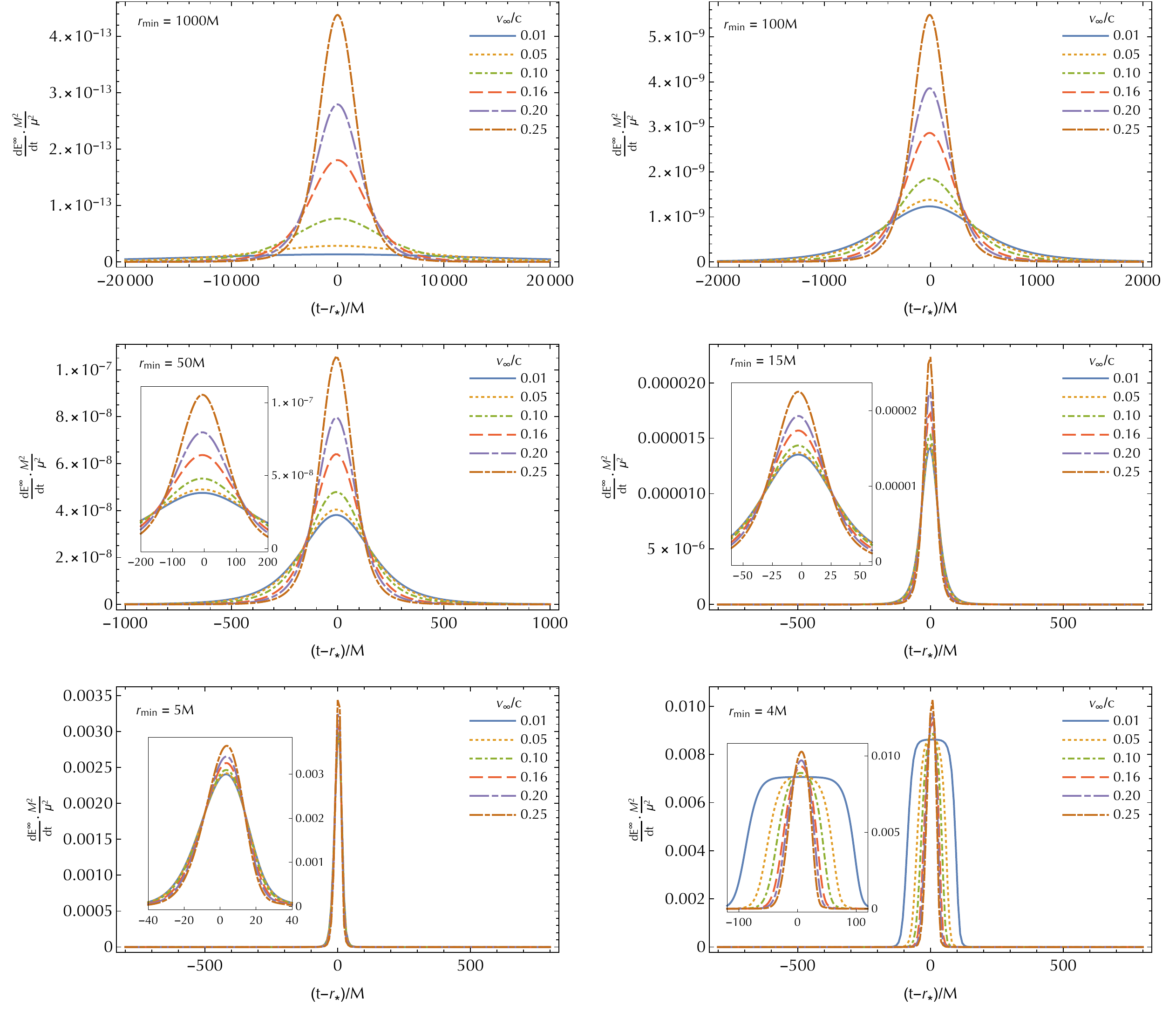}
\caption{\label{fig:fluxPlots}
Energy flux for a range of scattering trajectories. Note the extended
period of near-constant flux when $v_{\infty} = 0.01c$ and $r_{\rm min}=4M$.}
\end{figure*}
\begin{figure*}[th!]
\includegraphics[width=\textwidth]{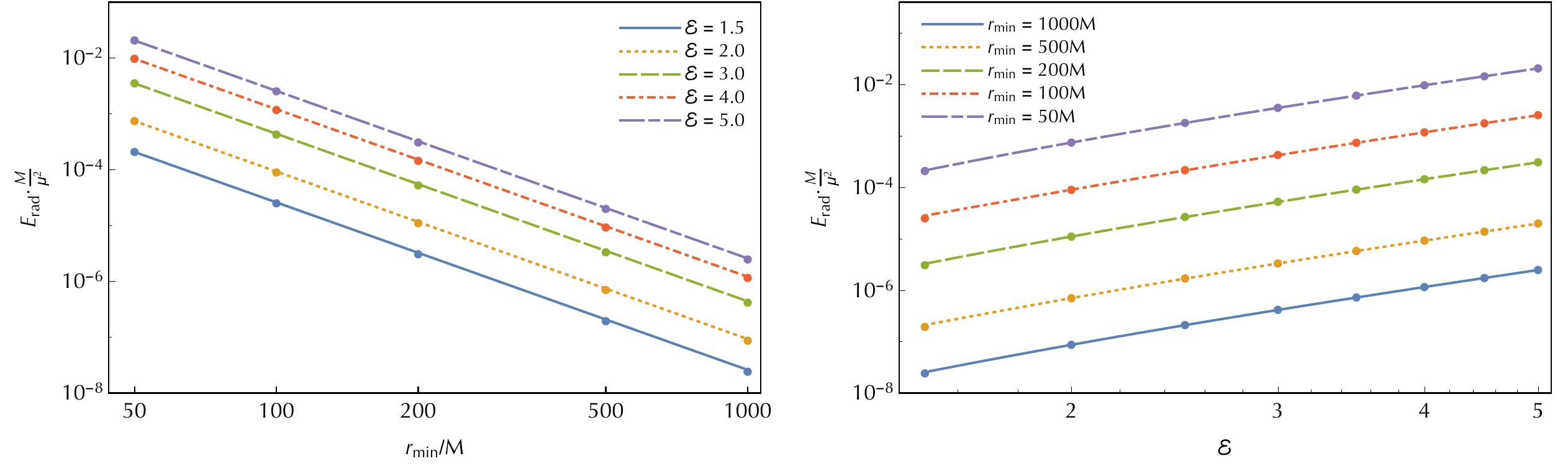}
\caption{\label{fig:totalRad}
Total radiation as a function of $r_{\rm min}$ (left) and $\mathcal{E}$ (right)
for our high-energy runs.
The lines connecting the dots are fits to our data as predicted by 
Eq.~\eqref{eq:PetersUltra}.}
\end{figure*}
%
\section{Gravitational waves from point-particle scattering}
\label{sec:results}

In this paper we are largely focused on the low-energy regime, and as such
the majority of our numerical runs (nearly 150 of them)
explore speeds of $0.1c \le v_\infty \le 0.25 c$ and   
periapses with $4M \le r_{\rm min}  \le 1000M$. We also performed 40 runs
with large energies $0.75c \lesssim v_\infty \lesssim 0.98 c$ for
periapses from $r_{\rm min} = 50M$ to $r_{\rm min} = 1000M$.
Our numerical results are summarized in Figs.~\ref{fig:fluxPlots}-\ref{fig:spectrumPlots}. 
Before examining the degree of numerical agreement between our 
results and various analytic predictions, we note some qualitative
features evident in our figures.

Emission of GWs is most effective for high-velocity, abruptly changing motion. Thus, high velocities and small periapses give rise to larger luminosities.
This is shown in Fig.~\ref{fig:fluxPlots}, where periapsis corresponds to $t-r_*=0$. Notice that the timescale over which emission occurs is not very sensitive on the initial velocity, but
depends mostly on the gravitational potential at periapsis. The luminosity changes by several orders of magnitude from $r_{\rm min}=1000M$ to $r_{\rm min}=4M$. 
Close to $r=4M$ new features set in. For example, the timescale for energy emission increases for small velocities. This feature is related solely to geodesic motion:
For small velocities, $r=4M$ is the capture threshold for incoming particles. Thus, the (point) particle can perform a large number of orbits before being scattered. Essentially then,
the flux is dictated by the circular orbit of similar radius. This property was also observed for plunges with large angular momentum~\cite{Berti:2010ce}, and has a visible impact on the waveform, as we show in Fig.~\ref{fig:waveformPlots},
and the spectrum in Fig.~\ref{fig:spectrumPlots} (see bottom panels). These
features are discussed in further detail below in Sec.~\ref{sec:smallLargeRMin}.

We now turn attention to the numerous 
analytical predictions that have been made for scattering
events over the years. 
In the following subsections we compare our numerical results with 
weak-field, post-Newtonian (PN) and zero-frequency limit 
(ZFL) predictions.

\subsection{Weak-field predictions}
\label{sec:weakField}
Peters \cite{Peters:1970mx} made a number of weak-field
predictions, two of which concern us here. His results
are valid when deflection angles are small and
velocities are constant.

First, in the limit of small velocities he obtains
\be
\frac{E_{\rm rad}}{M}=\frac{37\pi}{15}\frac{G^3}{c^5}\left(\frac{\mu}{M}\right)^2\frac{v}{(r_{\rm min}/M)^3}.\label{PertersE}
\ee
From a numerical comparison standpoint, the low-and-constant velocity
regime is challenging to explore since the particle
is always sped up as it approaches the BH. Our most-appropriate
run is $r_{\rm min} = 1000M$, $v_{\infty} = 0.25c$. 
While this trajectory is not particularly slow, it is nearly
constant-speed ($v_{\rm max} = 0.253c$) with small 
deflection angle, $1.9^\circ$.
In this case, our numerical 
value of $E_{\rm rad}=2.14\times 10^{-9} \mu^2/M$ is
within $10\%$ of Peters' prediction in Eq.~\eqref{PertersE}.
Meanwhile an event with the same
$r_{\rm min} = 1000M$, but $v_{\infty} = 0.01c$ deflects by an angle $131^\circ$
and has a maximum speed of $v_{\rm max} = 0.045c$. Unsurprisingly, in this case, we disagree
with Peters' prediction by an order of magnitude.

Peters also explores the high-energy limit, where he finds
the order-of-magnitude estimate
\be
\label{eq:PetersUltra}
\frac{E_{\rm rad}}{M}\sim \frac{G^3}{c^4}\left(\frac{\mu}{M}\right)^2\frac{{\cal E}^3}{(r_{\rm min}/M)^3}\,.
\ee
Constant velocities and high-energy are a natural fit, and
therefore we are able to explore the ultrarelativistic regime more thoroughly,
with results shown in Fig.~\ref{fig:totalRad}.
All the high-energy runs we consider have nearly constant speeds (within 1\%).
Deflection angles range between $6.6^\circ$ ($r_{\rm min} = 50M$, ${\cal E} = 1.5$) 
and $0.2^\circ$  ($r_{\rm min} = 1000M$, ${\cal E} = 5$). 
We first check the $(r_{\rm min}/M)^{-3}$ scaling of radiation, 
as shown in Eq.~\eqref{eq:PetersUltra}.
In the left panel of Fig.~\ref{fig:totalRad} we fit  $(r_{\rm min}/M)^{-3}$
lines to the data for several energies and see the expected behavior.
Our results are also consistent with an ${\cal E}^3$ dependence of the total radiated 
energy, at large and constant $r_{\rm min}$. 
This can be seen in the right panel of the same figure.
For several values of $r_{\rm min}$ we fit a function of the form 
$(A \mathcal{E}^3 + B \mathcal{E}^2 + C \mathcal{E}) \cdot (r_{\rm min}/M)^{-3}$.
While there is variation in the values of the $B$ and $C$ terms, we find consistent
leading-order behavior and are able to predict that the missing coefficient 
in Eq.~\eqref{eq:PetersUltra} is $28 \pm 2$.

\subsection{Post-Newtonian expressions}
%
\begin{figure*}[ht]
\includegraphics[width=\textwidth]{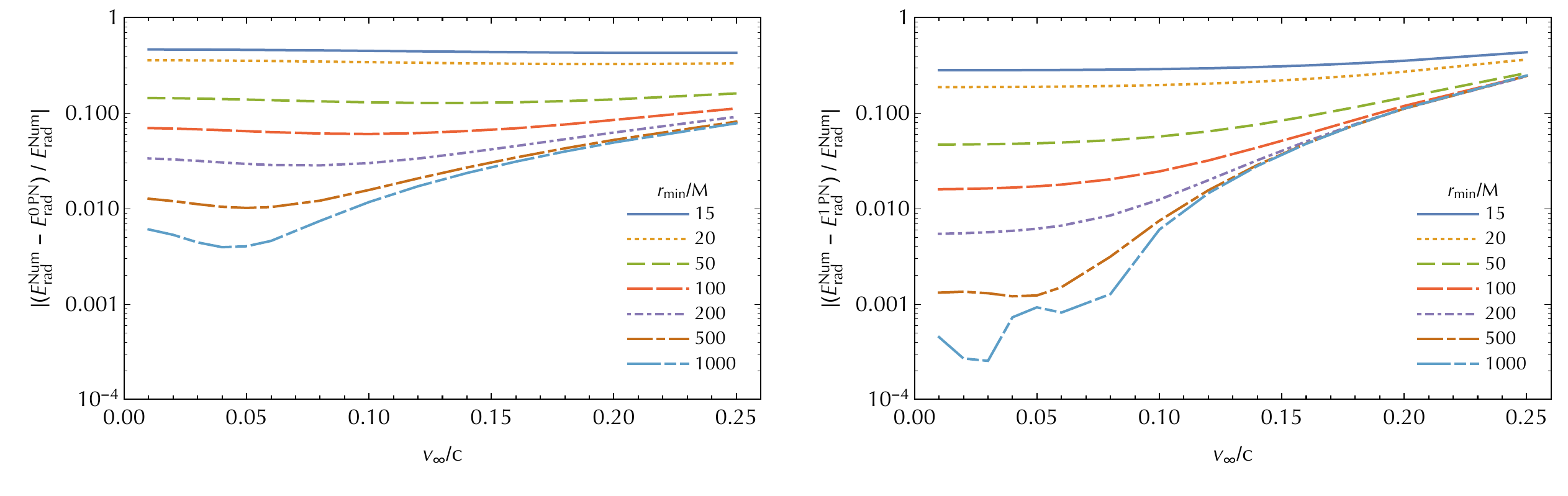}
\caption{
\label{fig:PNError}
The relative error between our numerical values and PN predictions, Eq.~\eqref{eq:PNFlux}, of the radiated energy for a range of scattering runs.
For low velocities we see the expected convergence with PN order, but at higher speeds (where the approximation $v_{\infty}^2 \sim G M/r_{\rm min}$
breaks down severely) the 1PN term actually worsens agreement.
The unevenness of the $r_{\rm min} = 1000M$ line in the right panel
likely indicates that we have reached the floor of our numerical accuracy.}
\end{figure*}
In a PN expansion, the radiated energy from a scattering event can be written in form
\begin{align}
\label{eq:PNFlux}
\frac{E_{\rm rad}}{M} =
\frac{2}{15 c^5} \frac{G^7}{(\mathcal{L}/M)^7} \frac{\mu^2}{M^2}  
\left( \mathcal{F}_0 + \frac{1}{c^2} \mathcal{F}_1  + \cdots \right),
\end{align}
where
\begin{align}
\begin{split}
&\mathcal{F}_0 = \Big( 96 + 292 e_r^2 + 37 e_r^4 \Big) \arccos (-1 / e_r) \\
& \hspace{5ex}
+
\frac{\sqrt{e_r^2 - 1}}{3}  (602 + 673 e_r^2), 
\end{split} \\
&\mathcal{F}_1  =
\frac{G^2}{(\mathcal{L}/ M)^2}\bigg[
\frac{1}{56} \arccos (-1/e_r) \notag
\\
& \hspace{4ex} \times\Big( 52624 + 117288 e_r^2 + 94542 e_r^4 + 17933 e_r^6 \Big)  \\
&+ 
\frac{\sqrt{e_r^2 - 1}}{840}
\Big( 1 516 596 + 1447788 e_r^2 + 1271421 e_r^4 \Big)
\bigg],
\notag
\end{align}
and 
\begin{align}
\begin{split}
e_r^2 =& 1 + 2 (\mathcal{E} - 1) \frac{\mathcal{L}^2}{G^2 M^2}\\
&+  \frac{\mathcal{E} - 1}{c^2} 
\left[ - 12 - 15 (\mathcal{E} - 1) \frac{\mathcal{L}^2}{G^2 M^2} \right]\,.	
\end{split}
\end{align}
As we are only interested in the point-particle limit, we have dropped all
higher-order terms in $\mu / M$ above.
The lowest order term ($\mathcal{F}_0$) is the scattering 
equivalent of the Peters-Mathews result~\cite{Peters:1963ux},
derived correctly first by Turner \cite{Turn77} 
(see also previous work \cite{Hansen:1972jt}).
Subsequently, Blanchet and and Sch\"afer \cite{Blanchet:1989cu} 
found the next-to-leading order term
$\mathcal{F}_1$. They used the quasi-Keplerian formalism 
\cite{DamoDeru85} wherein the Newtonian eccentricity `splits' into three
eccentricities after leading order. The above expressions use the `$r$-eccentricity'.

The results of our comparison with these PN predictions are summarized
in Fig.~\ref{fig:PNError}, with the 0PN residual on the left and
1PN residual on the right. We find expected order-of-magnitude agreement
for low velocities, but around $v_{\infty} = 0.1c$ 
the agreement starts to fail. Once we reach $v_{\infty} = 0.25c$,
the 1PN term actually \emph{worsens} the agreement for all $r_{\rm min}$. 
At first glance
this is a very puzzling result. However, we believe these
features are correct, as a comparison with the bound, eccentric
case makes clear.

The PN expansion of the orbit-averaged eccentric-motion flux
can be written in the form that is highly analogous
to Eq.~\eqref{eq:PNFlux} (see, e.g. \cite{Blanchet:2013haa}),
\begin{align}
\left\langle \frac{dE}{dt}\right\rangle = 
\frac{32 c^5}{5 G} \nu^2 y^5 \Big( \mathcal{I}_0 + y \mathcal{I}_1 
+ y^{3/2} \mathcal{I}_{3/2} + \cdots \Big) ,
\end{align}
where the ${\cal I}_i$ terms are `enhancement factors' akin to the
${\cal F}_{i}$ terms above.
The PN parameter $y \equiv \left( G M \Omega_\phi / c^3 \right)^{2/3}$ 
is a natural gauge invariant in which to expand, 
defined using the observable $\Omega_\phi$,
the average advance in the particle's azimuthal position.
It is of the same order as $v^2/c^2$ and $GM/(c^2 r)$.
Here $v$ and $r$ are the characteristic speed and separation
of the eccentric binary.

A scattering system does not exhibit such a clear PN parameter 
[hence, the counting of  PN orders with $c^{-2}$ in Eq.~\eqref{eq:PNFlux}]. 
The natural
length scale of the problem is $r_{\rm min}$, but since the system 
is unbound, we \emph{do not} 
find that $GM/r_{\rm min} \sim v_{\infty}^2$ in general. 
As can be seen in Fig.~\ref{fig:PNError}, 
this leads to less-than-uniform convergence in PN order
when we vary the speed of our scattering particle.
This can be contrasted with similar PN comparisons made for eccentric
motion, where PN-order convergence exists 
even for high eccentricities~\cite{ForsEvanHopp16}.
(We note that Turner and Will \cite{TurnWill78} attempted to address
this problem by including one order higher in the $v^2/c^2$ expansion
than in the $GM/(c^2 r)$ expansion. 
However, we find that their calculation neglects
$GM/(c^2 r)$ terms that are important and our agreement with their predictions is poor.
Hence, we do not compare with the Turner and Will result here.)

We note in passing that for bound motion the PN expansion of the energy flux
is known through 3PN \cite{ArunETC08b}, and to much higher order in the 
small mass-ratio limit \cite{ForsEvanHopp16}. Meanwhile, as far as we know,
the scattering expression has remained at 1PN for almost thirty years.
In principle, all the tools are available for an enterprising PN expert to 
extend the Blanchet and Sch\"afer \cite{Blanchet:1989cu} expression to 1.5PN and beyond.

Lastly, in addition to these PN results, 
Kovacs and Thorne~\cite{Kovacs:1978eu} provide a 
few more properties of the radiation emitted.
For example, they show (citing Ruffini and Wheeler \cite{RuffWhee71}) 
that the {\it energy spectrum} for 
low-energy scatterings behaves as
\be
\frac{dE}{d\omega}=\frac{32}{5}\left(\frac{\mu M}{b}\right)^2\left(\frac{\omega b}{v_\infty}\right)^3e^{-2\omega b/v_\infty}\,,
\label{kovacsthorne}
\ee
in the $\omega \gg v_\infty/b$ limit.
This is a prediction of the large-$\omega$ tail of the spectra,
which we plot in Fig.~\ref{fig:spectrumPlots}.
We fit this large-$\omega$ portion of the data from our runs to a function
of the form $A\omega^3 e^{-B \omega}$. When $r_{\rm min}$ is large, 
we find good agreement, even for moderately large velocities $v_\infty$. 
For example, for $r_{\rm min}=1000M,\,v_\infty=0.25c$, the best-fit 
values of $A$ and $B$ each agree with 
those of relation \eqref{kovacsthorne} to within $\sim 6\%$.

\begin{figure}
\includegraphics[width=\columnwidth]{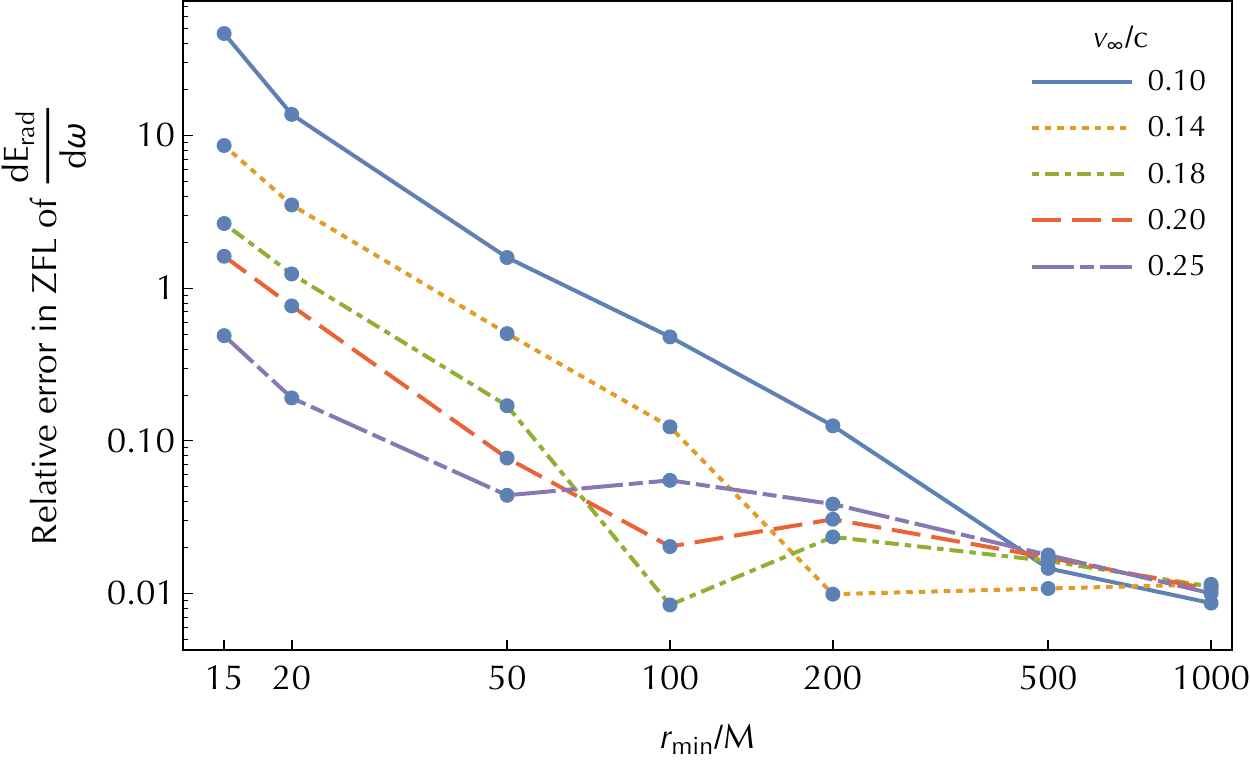}
\caption{\label{fig:ZFLError}
Error between our numerical calculation and Smarr's analytic
prediction, Eq.~\eqref{zfl}, of the energy radiated in the ZFL.
Our numerical ZFL can be seen graphically by looking at the $\omega = 0$ crossing of
the spectra in Fig.~\ref{fig:spectrumPlots}.
The unevenness of the lines near $r_{\rm min} \approx 100M$ 
is a result of zero crossings in the error.}
\end{figure}
%
\subsection{Zero-frequency limit}
The ZFL is a qualitatively interesting 
area of parameter space to explore.
Its relevance was pointed out by Weinberg in 1964, using quantum arguments~\cite{Weinberg:1964ew,Weinberg:1965nx}. 
Smarr first noticed that the ZFL can be applied successfully to classical problems of GW generation~\cite{Smarr:1977fy,Cardoso:2014uka}. Generically, the prediction
is that when two scattering bodies have non-zero speeds at infinite separation, their energy spectrum
does not vanish in the ZFL. Remarkable agreement with Smarr's calculation was found for point particles plunging into BHs~\cite{Cardoso:2002ay,Berti:2010ce}.
Surprisingly, results of full nonlinear calculations of head-on collisions of equal mass BHs at large center-of-mass energies
were in very good agreement with Smarr's linearized estimates~\cite{Sperhake:2008ga,East:2012mb,Cardoso:2014uka}.
For a scattering event, Smarr computes the ZFL of the energy spectrum to be
\begin{align}
\label{zfl}
&\left( \frac{dE}{d\omega} \right)_{\omega \to 0}
=\frac{4}{\pi} \frac{\mu^2 M^2 \mathcal{E}^2}{b^2}
\frac{(1+v^2)^2}{v^4}\\
& \hspace{10ex}
\times \left[2-\frac{16}{3} v^2+\left(3 v-\frac{1}{v}\right) \log \left(\frac{1+v}{1-v}\right)\right]\,.
\nonumber
\end{align}
His calculation assumes large impact parameters and constant velocities.
We expect these assumptions to be reasonably valid when $r_{\rm min} \gg 2M$,
and indeed, looking at Fig.~\ref{fig:ZFLError}, we find that for
$r_{\rm min} \gtrsim 200M$ our numerical values agree with Smarr's prediction
to at least 10\%. When $r_{\rm min} = 1000M$ our relative errors are on the order
of 1\% for all velocities.
\begin{figure*}
\includegraphics[width=\textwidth]{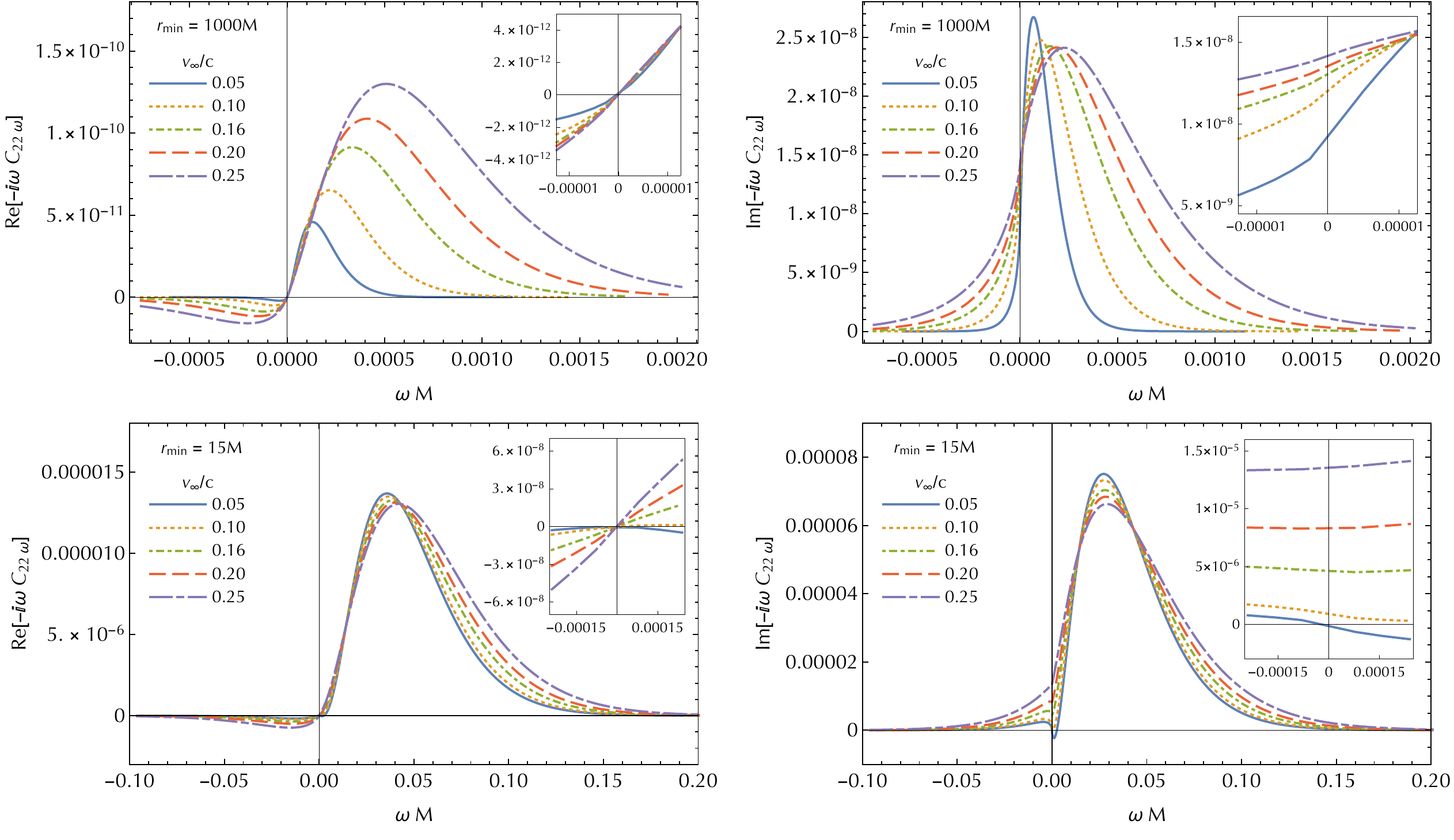}
\caption{\label{fig:omCoeffPlots}
The $(\ell, m) = (2,2)$ contribution to the
FD waveform (multiplied by $-i \omega$) for a range of scattering events
with $r_{\rm min} = 15M$ and $r_{\rm min} = 1000M$. 
The spectrum is sampled everywhere (except exactly
$\omega = 0$) with a step size $M\Delta \omega = 8 \times 10^{-5}$ 
(for $r_{\rm min} = 15M$) and $M\Delta \omega = 2.5 \times 10^{-6} $ 
(for $r_{\rm min} = 1000M$).
The modes are combined to form the TD waveform shown in Fig.~\ref{fig:waveformPlots},
using Eq.~\eqref{eq:waveform}.
The jump in that TD waveform between $u = -\infty$ and $u = \infty$ is 
predicted by the ZFL here, as shown in the insets.}
\end{figure*}
\begin{figure*}
\includegraphics[width=\textwidth]{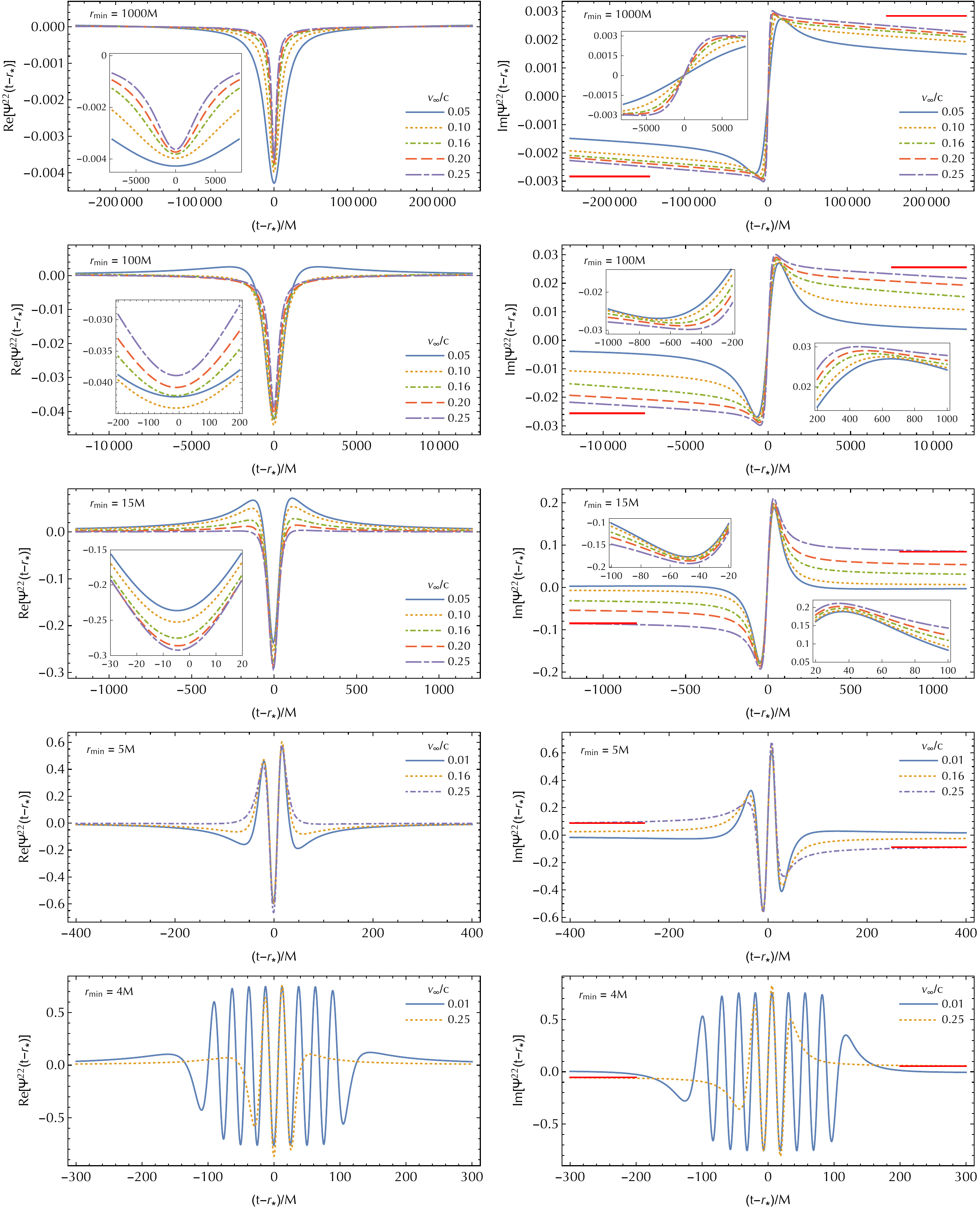}
\caption{\label{fig:waveformPlots}
The $(\ell, m) = (2,2)$ contribution to the waveform for
scattering events with a range of $r_{\rm min}$ and particle speeds. 
In the right panels
a solid red line shows the prediction of the memory effect
for $v_{\infty}/c = 0.25$ based on the
ZFL of the FD waveform shown in Fig.~\ref{fig:omCoeffPlots}.}
\end{figure*}

In addition to Smarr's prediction of the ZFL of the energy spectrum, 
we can use the ZFL to evaluate the difference in $\Psi_{\ell m}$ before and after the
encounter. Taking ZFL of the Fourier transform 
of $\pa_u \Psi_{\ell m} (u,r_*)$ (and performing an integration by parts) we have
\begin{align}
\begin{split}
&\lim_{\omega \to 0} \int_{-\infty}^\infty e^{i \omega u} \pa_u
\Psi_{\ell m} (u, r_*) du \\
&\hspace{10ex}
= \Psi_{\ell m} (\infty, r_*) - \Psi_{\ell m} (-\infty, r_*)\,.
\label{eq:ZFL}
\end{split}
\end{align}
We define the jump in $\Psi_{\ell m}$ as measured at infinity to be
$
\left\llbracket \Psi_{\ell m} \right\rrbracket \equiv
\Psi_{\ell m} (\infty, r_* \to \infty) - \Psi_{\ell m} (-\infty, r_* \to \infty).
$
It can be evaluated by combining Eqs.~\eqref{eq:ZFL} and \eqref{eq:waveform},
\begin{align}
\left\llbracket \Psi_{\ell m} \right\rrbracket &= \lim_{\omega \to 0} \int_{-\infty}^\infty e^{i \omega u} \pa_u
\left( \frac{1}{2\pi} \int_{-\infty}^\infty C^+_{\ell m\o'} \, e^{-i \o' u}  d\o' \right) du \nonumber\\
&= - \lim_{\omega \to 0} i \omega \, C^+_{\ell m \omega}\,.\label{zflim}
\end{align}
In Fig.~\ref{fig:omCoeffPlots} we show plots of $- i \omega C^+_{\ell m \omega}$,
with insets showing the $\omega \approx 0$ regime.
As seen in that figure, the ZFL only has an imaginary component.
For trajectories that travel close to the BH, 
it yields precisely the expected offset
in the asymptotic TD waveforms, as shown in Fig.~\ref{fig:waveformPlots}. 
We see that there is a very good agreement for values of $r_{\rm min} \lesssim 15M$.
The lack of agreement for runs with larger periapses is discussed in the next section.
The offset in the TD waveform is a well-known phenomenon also called 
the memory effect \cite{ZeldPoln74,Chri91}.  
See Favata's review \cite{Fava10} and references therein for 
a thorough discussion of the subject. 

\subsection{Trajectories with large and small values of $r_{\rm min}$}
\label{sec:smallLargeRMin}

From a qualitative standpoint, the most interesting areas of parameter
space we explore are those with very small and very large
pericenters. We now consider the features 
of these two regimes each in turn, starting with close encounters.

The trajectory $r_{\rm min} = 4M, {\cal E} = 1$ is 
marginally bound and parabolic. 
A particle on this trajectory will
orbit the BH an infinite number of times and radiate
an infinite amount of energy (representing, of course,
a breakdown in the geodesic approximation). We approach
this point considering scattering events with $v_{\infty}$ as
small as $0.1 c, ({\cal E} \approx 1.00005)$. In this case there is a clear qualitative shift
in the spectrum, energy flux, and waveform relative to the 
other cases we consider. 

\begin{figure*}
\includegraphics[width=\textwidth]{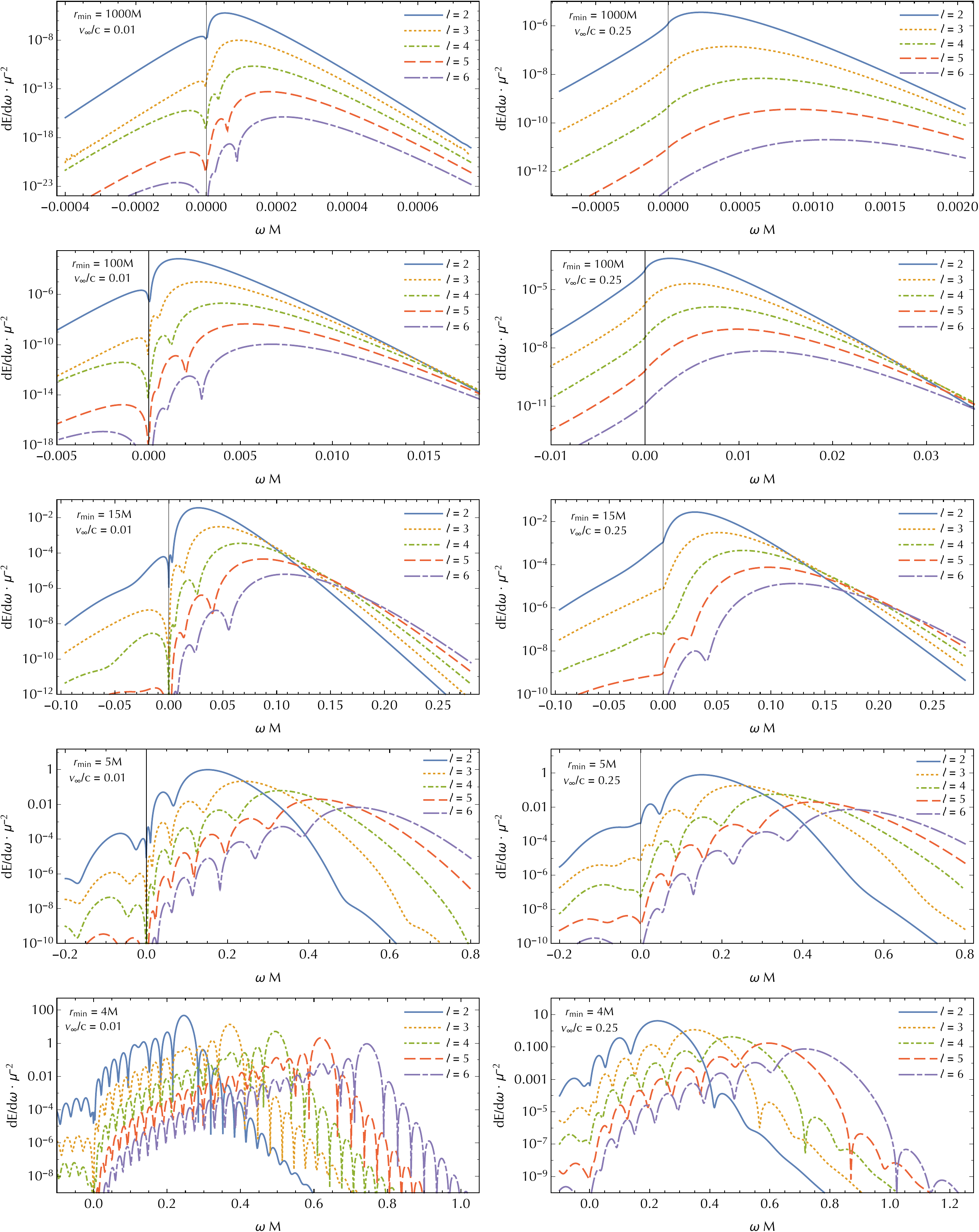}
\caption{
\label{fig:spectrumPlots}
The radiated energy spectrum for runs with $v_{\infty}/c =0.01$ and $0.25$
and a range of $r_{\rm min}$.
The top and bottom rows show spectra for the largest 
and smallest periapses considered. In those cases, there is a 
sizable effect from due to the value of $v_{\infty}$.
This is most evident in the differences in axes of the plots 
in the left and right columns.
Meanwhile, for moderate $r_{\rm min} = 15M$, the trajectories
with $v_{\infty}/c =0.01$ and $0.25$ yield spectra 
that are qualitatively quite similar.}
\end{figure*}

Consider first the spectrum shown in the bottom left of 
Fig.~\ref{fig:spectrumPlots}. The spectrum of each $\ell$
mode is dominated by the $\ell = m$ contribution.
The peak for each $\ell$ mode
occurs at $\omega = m \Omega_{4M} = m/(8M)$, the $\ell = m$ harmonic 
of the fundamental frequency of a circular orbit at $r_p = 4M$.
The waveform for this trajectory (bottom row of Fig.~\ref{fig:waveformPlots})
and the flux (bottom right of Fig.~\ref{fig:fluxPlots}) show evidence
of the particle zoom-whirling close to the BH. In fact,
when $v_{\infty}=0.01c$, the particle remains close to the 
BH emitting a constant flux for nearly $200M$. 

The $r_{\rm min} = 4M$ runs also allow us to examine the 
conjectured Dyson-Gibbons-Thorne bound of a peak luminosity of
(restoring physical units) $c^5/G$. Presumably larger luminosities are impossible to achieve
since the radiation itself would then collapse to form a BH.
The bottom right panel of Fig.~\ref{fig:fluxPlots} shows that 
our flux peaks around $dE/dt\sim 0.01 \mu^2/M^2$ when $v_{\infty} = 0.25 c$.
Thus, even when extrapolating our results to equal-mass scatters,
the peak luminosity is below unity. This is an interesting result, 
indicating that the conjectured Dyson-Gibbons-Thorne bound holds~\cite{Cardoso:2013krh}.
Trajectories with $v_{\infty} > 0$ can penetrate the $r_{\rm min} = 4M$ boundary,
with the limit approaching $r_{\rm min} = 3M$ as ${\cal E} \to \infty$ 
($v_{\infty} \to c$). In a future work, we will explore how close
to $r_{\rm min} = 3M$ our code can reach, and, as a result, see how 
close to the conjectured bound these ultrarelativistic encounters
bring us.

At the other extreme are runs where $r_{\rm min}$ gets very large.
In this work we consider periapses as large as $1000M$. 
We have seen that the radiated energy trends as $r_{\rm min}^{-3}$,
and so these weak field scatters radiate very weakly. Indeed,
we see in Fig.~\ref{fig:spectrumPlots} that their spectrum
is peaked around very small frequencies. Numerically these
frequencies are quite challenging for our machine-precision code.
In practice our GSL~\cite{GSL} integrator fails below 
$\omega_{\rm min} = 2.5 \times 10^{-6} / M$. This smallest frequency
provides a fundamental limit to our spectral method. It implies that 
any TD signal we reproduce will be periodic over a timescale of
$2 \pi / \omega_{\rm min} \approx 2.5 \times 10^6 M$.
This effect is plainly visible in the waveforms at the top of
Fig.~\ref{fig:waveformPlots} and it is the source of our
disagreement with the predicted memory effect. 
Indeed all of our waveforms eventually
repeat, but the beginning of this effect is only visible 
in the top two rows of waveforms, which are plotted over very long
timescales. We expect that if our machine-precision code could 
reach arbitrarily-small frequencies, we would see the exact memory 
effect predicted by the ZFL.
%

\section{Discussion}
\label{sec:conc}
Our results are in agreement with a number of approximations made in the literature, mostly for small velocities and large impact-parameter scatters.
We find strong evidence that the PN approximation is working well and converging in the regime where it should (low velocities).
This study is a first step in the broader program of understanding gravitational radiation from bound and unbound motion.
Left for future work is the study of high-energy scatters and plunges, and how they impact on peak luminosities (and consequences for the conjectured bound on luminosity) and other radiation properties. These questions may have some relevance for astrophysics, but they certainly have a bearing on our understanding of gravity at low- and high-energy scales.
\begin{acknowledgments}
We thank Luc Blanchet for useful correspondence.
The authors acknowledge financial support provided under the European Union's H2020 ERC Consolidator Grant ``Matter and strong-field gravity: New frontiers in Einstein's theory'' grant agreement no. MaGRaTh--646597.
Research at Perimeter Institute is supported by the Government of Canada through Industry Canada and by the Province of Ontario through the Ministry of Economic Development $\&$
Innovation.
This article is based upon work from COST Action CA16104 ``GWverse'', supported by COST (European Cooperation in Science and Technology).
This work was partially supported by the H2020-MSCA-RISE-2015 Grant No. StronGrHEP-690904..
The authors thankfully acknowledge the computer resources, technical expertise and assistance provided by CENTRA/IST. Computations were performed at the cluster
``Baltasar-Sete-S\'ois,'' and supported by the MaGRaTh--646597 ERC Consolidator Grant.
\end{acknowledgments}

\bibliography{ScatteringPN}

\end{document}